\documentclass[conference]{IEEEtran}
\IEEEoverridecommandlockouts

\usepackage{cite}
\usepackage{amsmath,amssymb,amsfonts}
\usepackage{algorithmic}
\usepackage{graphicx}
\usepackage{textcomp}
\usepackage{xcolor}
\usepackage{color}
\usepackage{url}
\usepackage{hyperref}

\usepackage{diagbox}    

\usepackage[latin9]{inputenc}
\usepackage{epstopdf}
\usepackage{amsthm}
\usepackage{amsmath}
\usepackage{amsthm}
\usepackage{amssymb}
\usepackage{graphicx}
\usepackage{subfigure}
\usepackage{wrapfig}
\usepackage{algorithmic}
\usepackage{algorithm}
\usepackage{mathrsfs}
\usepackage{float}
\usepackage{mathtools}
\usepackage{tabulary}
\usepackage{booktabs}
\usepackage{caption}
\usepackage{multirow}
\usepackage{array}
\usepackage{makecell}
\usepackage{graphicx}
\usepackage{enumitem}
\usepackage{subcaption}
\usepackage{dirtree}

\usepackage{titlesec}
\titlespacing*{\section}      {0pt}{*1.0}{*0.25}  
\titlespacing*{\subsection}   {0pt}{*0.25}{*0.25}   
\titlespacing*{\subsubsection}{0pt}{*0.35}{*0.1}   

\def\BibTeX{{\rm B\kern-.05em{\sc i\kern-.025em b}\kern-.08em
    T\kern-.1667em\lower.7ex\hbox{E}\kern-.125emX}}

\begin{document}







\title{Testbed and Dataset Collection of Bluetooth RF Fingerprints for Deep Learning-Based Device Identification and Beyond}

\title{Dataset and Experimental Testbed for Deep Learning-Based Bluetooth Device Fingerprinting}


\author{Haytham Albousayri and Bechir Hamdaoui\\
\small School of Electrical Engineering and Computer Science, Oregon State University, Corvallis, OR, USA\\
Emails: \{albousah, hamdaoui\}@oregonstate.edu 
\thanks{This work is supported in part by NSF Award No. 2350214.}}

\maketitle

\thispagestyle{empty}
\pagestyle{empty}

\begin{abstract}
AI-enabled wireless networks have emerged as vital components of future-generation wireless communication systems. Among various wireless architectures, Bluetooth Low Energy (BLE) networks have attracted significant research interest due to their widespread adoption and low cost. In these networks, deep learning (DL) models can be integrated directly into the physical-layer infrastructure to perform diverse tasks, including environment sensing, localization, radio frequency fingerprinting (RFFP), and signal denoising.
While highly effective under their original training configurations, these DL models often fail significantly when deployed in unseen environments---such as different rooms, varying temperature conditions, or altered carrier frequencies. Addressing this limitation requires rigorous benchmarking of models and algorithms against unseen configurations to evaluate their generalization and robustness.
To bridge this gap, this paper makes two key contributions: \textit{i)} we introduce a comprehensive, large-scale wireless BLE dataset consisting of raw IQ transmissions from 31 distinct commercial devices captured across diverse days, locations, environments, temperatures, receiver hardware, and frequency channels; \textit{ii)} we present a case study evaluating state-of-the-art techniques for domain-robust RFFP to advance real-world AI reliability.
\end{abstract}


\section{Introduction}
\label{sec:introduction}
As wireless research shifts away from traditional model-based paradigms toward data-driven solutions, deep learning (DL) has emerged as a cornerstone for signal analysis, channel estimation, and physical-layer intelligence. While offering very good precision, these DL models tend to suffer from a major drawback: a tendency to overfit to their specific training environments, severely limiting their generalization to unseen domains~\cite{johnson2025domain, jagannath2023embedding, tang2024causal, xing2022design, elmaghbub2023eps, albousayri2025neural}.
To ensure the robustness of these DL models, evaluations must look beyond the training domain by rigorously testing them against unseen conditions such as different temperatures, locations, and multipath environments, carrier frequencies, and receiver hardware. 

In response to this need, this paper introduces a comprehensive, large-scale Radio Frequency (RF) dataset consisting of massive signal data captured from 31 distinct commercial transmitting Bluetooth devices using two Ettus USRP B210 software-defined radios (SDRs). The released dataset provides the wireless research community with a first-of-its-kind, multi-domain experimental benchmark designed specifically to study the effectiveness, resilience, and generalization capabilities of data-driven wireless techniques, spanning various AI-driven wireless applications including deep learning-based integrated sensing and communication (ISAC) and physical-layer RF fingerprinting (RFFP). By accounting for diverse, real-world domain shifts---including operational temperature, wireless channel, physical environments, and receiver hardware heterogeneity---the collected dataset provides critical insights into how joint and individual variations affect data-driven frameworks in the Internet of Things (IoT) ecosystem.

Additionally, we study the impact of domain adaptation in RFFP. Specifically, we experimentally evaluate how changing transmitter locations and receiver hardware affects the performance of DL-based device identification systems. Further, our study provides a comprehensive evaluation against state-of-the-art DL architectures and signal processing benchmarks.

\subsection{What Distinguishes Our Dataset From Existing Ones?}
\begin{table*}[t]
\centering
\caption{Comparison of publicly available Bluetooth RF fingerprinting datasets.}
\label{tab:dataset_comparison}
\setlength{\tabcolsep}{5pt} 
\renewcommand{\arraystretch}{0.7}
\begin{tabular}{lccccccccccc}
\toprule
\multirow{2}{*}{\textbf{Dataset}} &
\multirow{2.5}{*}{\textbf{\# Devices}} &
\multirow{2.5}{*}{\makecell{\textbf{Frames}\\\textbf{Extracted}}} &
\multirow{2.5}{*}{\makecell{\textbf{Features}}}  &
\multirow{2.5}{*}{\makecell{\textbf{Publicly}\\\textbf{Available}}}  &
\multicolumn{6}{c}{\textbf{Domain Shifts}} \\

\cmidrule(lr){6-11}

&&&&&
\textbf{\scriptsize Location}\tiny &
\textbf{\scriptsize Time} \tiny&
\textbf{\scriptsize Environment} \tiny&
\textbf{\scriptsize Temperature} \tiny&
\textbf{\scriptsize Receiver (\#)} &
\textbf{\scriptsize Channel} \\

\midrule

\scriptsize Uzundurukan \emph{et al.} (2020) \cite{uzundurukan2020database}
& 20 
& \checkmark
& IQ
& \checkmark
& \checkmark
& $\times$
& $\times$
& $\times$
& $\times$ (1)
& $\times$
\\

\scriptsize Jagannath \emph{et al.} (2023) \cite{jagannath2023embedding}
& 10 
& $\times$
& IQ
& \checkmark
& \checkmark
& \checkmark
& $\times$
& $\times$
& $\times$ (1)
& $\times$
\\

\scriptsize Kashani  \emph{et al.} (2024) \cite{kashani2024radio}
& 13 
& \checkmark
& IQ
& \checkmark
& \checkmark
& \checkmark
& $\times$
& $\times$
& $\times$ (1)
& \checkmark
\\

\scriptsize Pereira  \emph{et al.}(2026) \cite{1s6a-tv49-26}
& 4 
& \checkmark
& IQ
& \checkmark

& \checkmark
& \checkmark
& $\times$
& $\times$
& $\checkmark$ (4)
& \checkmark
\\

\scriptsize Monteiro  \emph{et al.} (2026) \cite{monteiro2026deep}
& 1 
& NA
& RSSI
& \checkmark

& \checkmark
& \checkmark
& $\times$
& $\times$
& $\checkmark$ (6)
& $\times$
\\

\midrule
\textbf{\scriptsize Our Dataset}
& 31 
& \checkmark
& IQ
& \checkmark
& \checkmark
& \checkmark
& \checkmark
& \checkmark
& $\checkmark$ (2)
& \checkmark
\\

\bottomrule
\end{tabular}
\vspace{-10pt}
\end{table*}

While most open-source RF benchmarks focus on Wi-Fi or LoRa~\cite{elmaghbub2021comprehensive, hanna2022wisig}, publicly available Bluetooth datasets remain exceedingly limited. Existing open-source datasets either simply report high-level indicators such as Received Signal Strength Intensity (RSSI), or are restricted to a single experimental setup.
%
In the field of RFFP for example, Uzundurukan et al.~\cite{uzundurukan2020database} released sampled raw passband Bluetooth Classic signals but only at a static collection environment. To introduce spatial variability, Jagannath et al.~\cite{jagannath2023embedding} captured raw IQ-based unframed horizon across varied transmission distances and line-of-sight (LoS) configurations; however, their work omitted receiver-side variations and internal hardware thermal fluctuations.
Kashani et al.~\cite{kashani2024radio} provided a framed IQ-based dataset.
While their study captured signals across distinct physical locations, the evaluations remained constrained to a single macro-environment and a uniform receiver architecture. 
Parallel to these security datasets, recent BLE benchmarks target ISAC tasks. For example, Pereira et al.~\cite{1s6a-tv49-26} use IQ-based features for human pose estimation, while Monteiro et al.~\cite{monteiro2026deep} utilize shallow RSSI for localization. 



Ultimately, most of these prior works suffer from a small number of transmitting devices and highly restricted deployment setups, making it difficult to evaluate the scalability and real-world generalization of their trained model.
Our work bridges these distinct gaps by introducing a comprehensive, publicly available BLE dataset that systematically spans extensive over-the-air channel configurations and hardware states. The proposed dataset captures transmissions across a dual-receiver framework, four distinct BLE data channels, and multiple physical environments under both wired and wireless deployment configurations. Crucially, our dataset explicitly logs high-fidelity signals synchronized with real-time temperature telemetry during active hardware warming-up phases, providing the research community with the multi-domain variation necessary to evaluate robust wireless machine learning models.
Table~\ref{tab:dataset_comparison} summarizes the existing datasets in terms of number of devices and deployed domains, highlighting the novelty and importance of our work. 



\subsection{Our BLE Dataset in Brief}
Our RF dataset provides high-fidelity, time-domain IQ samples collected using an experimental IoT testbed consisting of 31 distinct commercial BLE transmitting devices and two Ettus USRP B210 software-defined radio (SDR) receivers. The raw received signals are downconverted and digitized to capture fine-grained physical-layer features. The recorded data are stored as raw binary IQ frames. 

Our dataset consists of 16 distinct experimental scenarios (labeled D1 through D15, where D11 consists of D11-a/b), which are systematically summarized in Table~\ref{tab:dataset_summary} and can be downloaded at~\url{https://research.engr.oregonstate.edu/hamdaoui/datasets}. 
These setups are categorized into four scenarios:
\begin{itemize}[noitemsep, topsep=0pt, leftmargin=*]
    \item \textbf{Setup 1) Temperature-Aware Indoor Wireless (D1--D3):} This scenario captures the hardware warm-up phase of the devices, featuring active temperature telemetry logging. Transmissions are received from different locations: 1.5m LoS (D1), 1.5m NLoS (D2), and 7.5m LoS (D3).
    
    \item \textbf{Setup 2) Multi-Channel Indoor Wired (D4--D8):} This setup employs a coaxial wired connection in a stable thermal state. This setup spans four different RF channels (Ch1, 2, 14, and 32) using Rx1 (D4--D7), as well as a channel-matched receiver hardware shift using Rx2 (D8).
    
    \item \textbf{Setup 3) Multi-Receiver/Channel/Day Indoor Wireless (D9--D11):} This scenario introduces over-the-air channel and receiver hardware variations at a fixed distance of 1.5m LoS under stable thermal conditions. Data is captured across different channels (Ch1 vs. Ch2) and receiver front-ends (Rx1 vs. Rx2). This setup also includes two measurement runs collected on different days under identical physical configurations to evaluate cross-day model generalization.
    
    \item \textbf{Setup 4) Multi-Distance Outdoor Wireless (D12--D15):} Deployed in an outdoor environment, this setup captures transmissions across varying propagation distances (1m, 2m, 3m, and 1.5m LoS). 
\end{itemize}

The rest of the paper is organized as follows. Section~\ref{sec:testbed} describes the experimental testbed and hardware components. Section~\ref{sec:experiments} presents the dataset collection procedures. Section~\ref{sec:dataset} details the data preprocessing stage and stored format. Section~\ref{sec:applications} provides the results for a RFFP use case for the dataset. Section~\ref{sec:applications2} provides a discussion for other use cases. Section~\ref{sec:conclusion} concludes the paper.

\section{Testbed Description}
\label{sec:testbed}
This section introduces the BLE Physical-layer standard and details the hardware and software used to build our testbed.
\begin{figure}
    \centering
    \includegraphics[width=\linewidth]{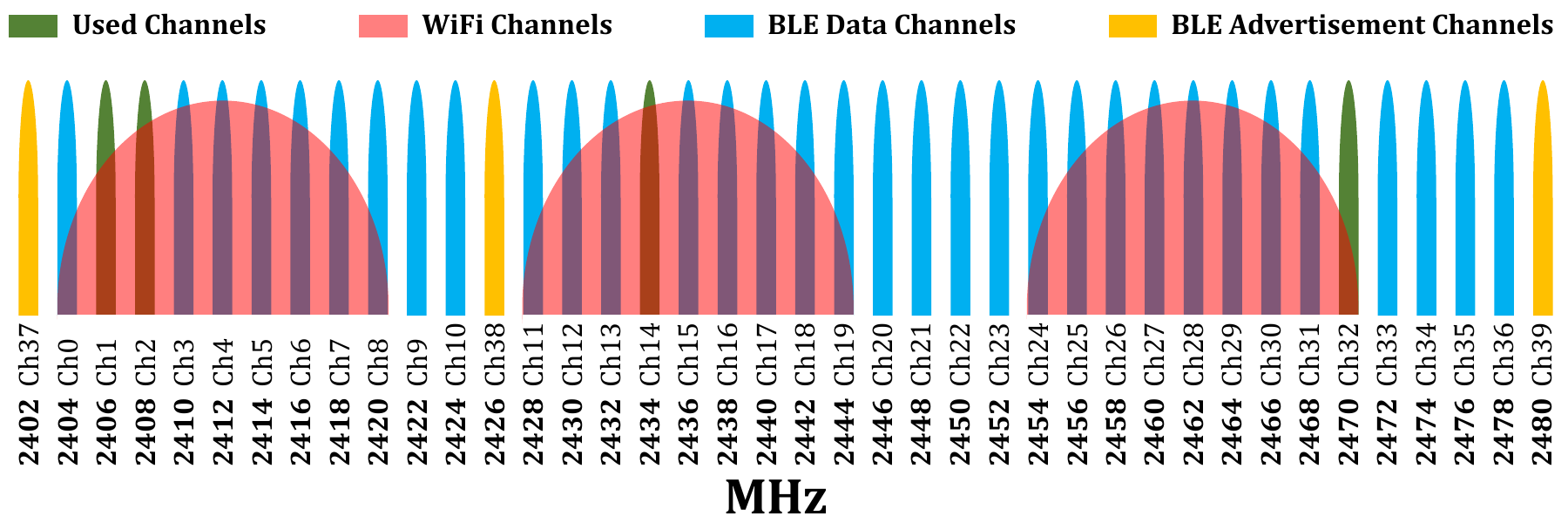}
    \caption{BLE channel map.}
    \label{fig:BLE_Channels}
\end{figure}
\subsection{Bluetooth Low Energy (BLE) Physical Layer Overview}
To contextualize the underlying signal characteristics captured in our dataset, we utilize the baseline LE 1M Uncoded PHY configuration as specified by the Bluetooth SIG. Operating within the $2.4\text{GHz}$ unlicensed ISM band ($2400$--$2483.5\text{MHz}$), BLE divides the spectrum into 40 RF channels spaced $2\text{MHz}$ apart, consisting of 3 primary advertising channels for discovery and 37 data channels for payload transfer~\cite{bluetooth60}, as shown in Figure~\ref{fig:BLE_Channels}.
Standard deployments employ a Frequency-Hopping Spread Spectrum (FHSS) mechanism to pseudo-randomly hop across data channels to mitigate interference and fading. The LE 1M PHY operates at a symbol rate of $1\text{Msym/s}$ to deliver a raw data rate of $1\text{Mbps}$ using Gaussian Frequency Shift Keying (GFSK) modulation with a nominal modulation index of $0.5$ and a bandwidth-bit period product (BT) of $0.5$~\cite{bluetooth60}. Over the air, each generated physical-layer packet contains a sequential structure starting with an 8-bit alternating preamble (\texttt{0xAA} or \texttt{0x55}) used by our preprocessing pipeline for arrival time synchronization, followed by a 32-bit Access Address for link filtering, a variable-length Protocol Data Unit (PDU) spanning 2 to 257 bytes of core payload, and a trailing 24-bit Cyclic Redundancy Check (CRC) for error detection.

\begin{figure}
    \centering
    \includegraphics[width=0.9\linewidth]{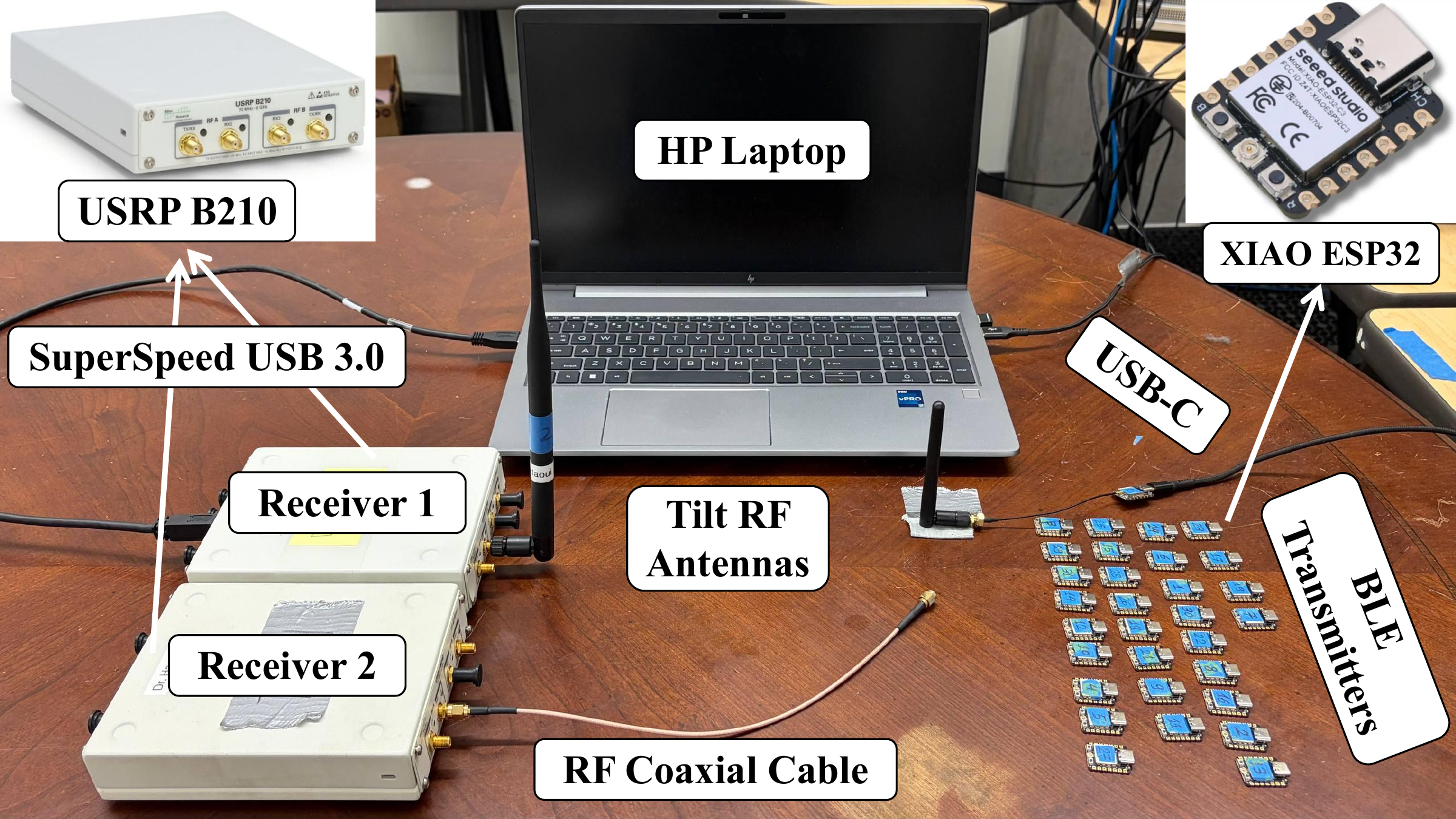}
    \caption{The BLE testbed consisting of 31 ESP32C3 transmitters and two USRP B210 receivers.}
    \label{fig:testbed}
\end{figure}

\subsection{Hardware Description} 
Here, we describe the hardware components of the testbed. As shown in Figure~\ref{fig:testbed}, we use Tilt RF antennas for wireless transmissions and coaxial cables for wired connections. Specifics regarding the transmitter and receiver hardware follow.
\subsubsection{Transmitter Side}

The testbed has 31 Seeed Studio XIAO IoT devices---mini-development boards based on the Espressif ESP32C3 WiFi/BLE dual-mode transceiver, programmed to send BLE traffic with random payload on the data channels. 
To study the frequency hopping impact on the fingerprint and isolate it from the other factors, we hardcoded these chips to send the signals over 4 BLE frequency channels, Channel 1 (Ch1), Channel 2 (Ch2), Channel 14 (Ch14) and Channel 32 (Ch32), where each one of these channels is centered at $2.406\text{GHz}$, $2.408\text{GHz}$, $2.434\text{GHz}$ and $2.470\text{GHz}$, respectively.

\subsubsection{Receiver Side}
\label{hw:rx}
Two Ettus Research USRP B210 software-defined radios are used to receive, down-convert and sample the signals coming from each of the 31 transmitters described above. While each transmitter periodically transmits traffic, the receiver continuously samples and records the baseband complex IQ signals across the entire time horizon. The front-end filter used by the receiver is set to $2\text{MHz}$ matching the bandwidth of the received signal, the sampling rate was set to $6\text{MS/s}$. 
The central frequency of the receiver is manually selected to match the carrier of the transmitter, for example $2.406\text{GHz}$ is picked when collecting the data over Ch1, $2.406\text{GHz}$, etc. 
\begin{figure}
    \centering
    \includegraphics[width=\linewidth]{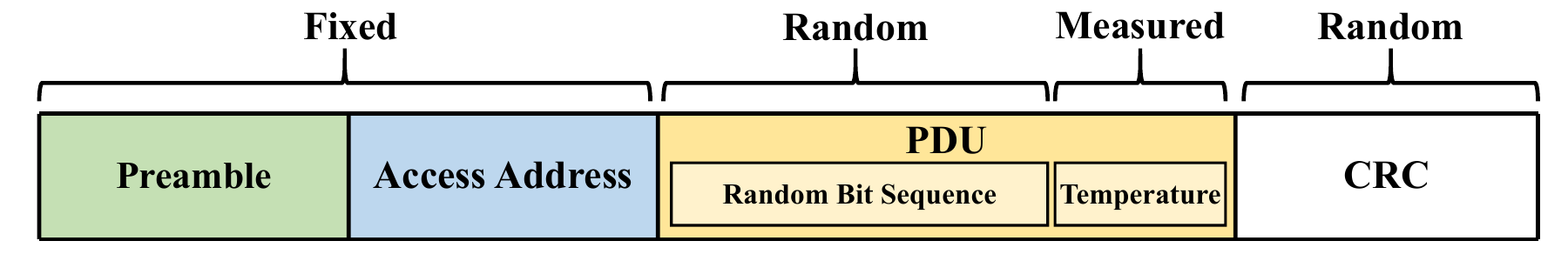}
    \caption{Designed BLE frame.}
    \label{fig:BLE_frame}
\end{figure}
\subsection{Software Description}
\subsubsection{Transmitter Side}
We configured our ESP32-C3 boards using C programming language, for that, we used ESP-IDF extension on Visual Studio. Besides using the basic setup and firmware to successfully build and run the firmware, our code implementation can be found here~\url{https://research.engr.oregonstate.edu/hamdaoui/datasets}, which has the following core design ideas:
\begin{itemize}[noitemsep, topsep=0pt, leftmargin=*]
    \item All devices were hardcoded with the same Physical address, that is \texttt{0x12:55:55:55:55:55}. This is to ensure the trainable classifier uses the hardware impairments to identify devices rather than overfitting to the unique address assigned to each device by default. 
    \item To be able to send connection-less period advertisement over the data channels (instead of the dedicated advertisement channels), we set up an extended advertisement mode, where such condition is allowed.
    Moreover, the default extended advertisement sends the traffic over all 37 data channels pseudo randomly; to control the transmission channel, we applied black listing to all channels except the target one.
    \item For experiments reporting temperature, an internal sensor measurement is injected into the last two bytes of the link layer payload (shown in Figure~\ref{fig:BLE_frame}), while the remaining bytes are configured to be either random or fixed. 
\end{itemize}



\subsubsection{Receiver Side}
We controlled the SDRs using an HP ZBook Power 15.6-inch G10 Mobile Workstation running Ubuntu Linux. This host computer is equipped with a 13th Gen Intel Core i7 processor and 64 GB DDR5-5600 RAM; we selected Linux because we found it offers significantly better performance for SDR operations than Windows.
To configure our USRP B210 SDR, we utilized GNU Radio, a Python-based graphical tool for real-time radio control and signal processing. After setting up the environment with the appropriate parameters (as detailed in Section~\ref{hw:rx}), we executed our flowgraph to capture time-domain IQ signals and save them locally for subsequent cleaning and processing. Additionally, we integrated real-time visual plots of the time-domain IQ samples, frequency spectrum, and waterfall graphs to facilitate active troubleshooting during data collection.

\subsection{Conventions and Notations}
Throughout this paper, the two receivers used are referred to as Rx1 and Rx2; IoT, device and transmitter (Tx) are used interchangeably; the term "channel" and BLE frequency channels are used interchangeably; the terms "signal" and "frame" are used interchangeably to refer to the sampled PHY layer baseband signal of BLE transmitters.  

\section{Data Collection and Experimental Scenarios}
\label{sec:experiments}


\begin{figure}
    \centering
    \includegraphics[width=0.75\linewidth]{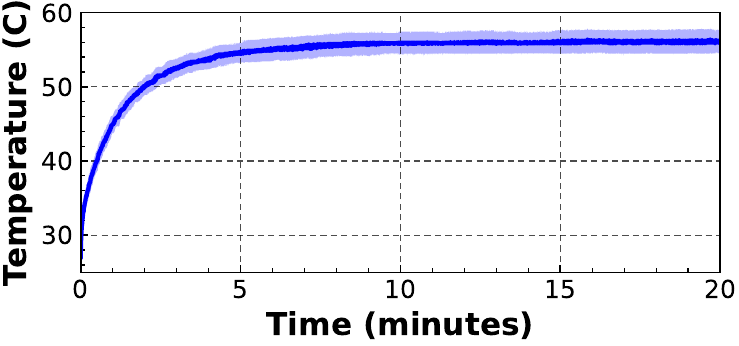}
    \caption{Device average temperature after powered-on~\cite{albousayri2026characterizing}.}
    \label{fig:TempvsTime_Workshop}
\end{figure}

\begin{figure*}[!t]
    \centering
\subfigure[1.5m LoS]{\includegraphics[width=0.144\linewidth]{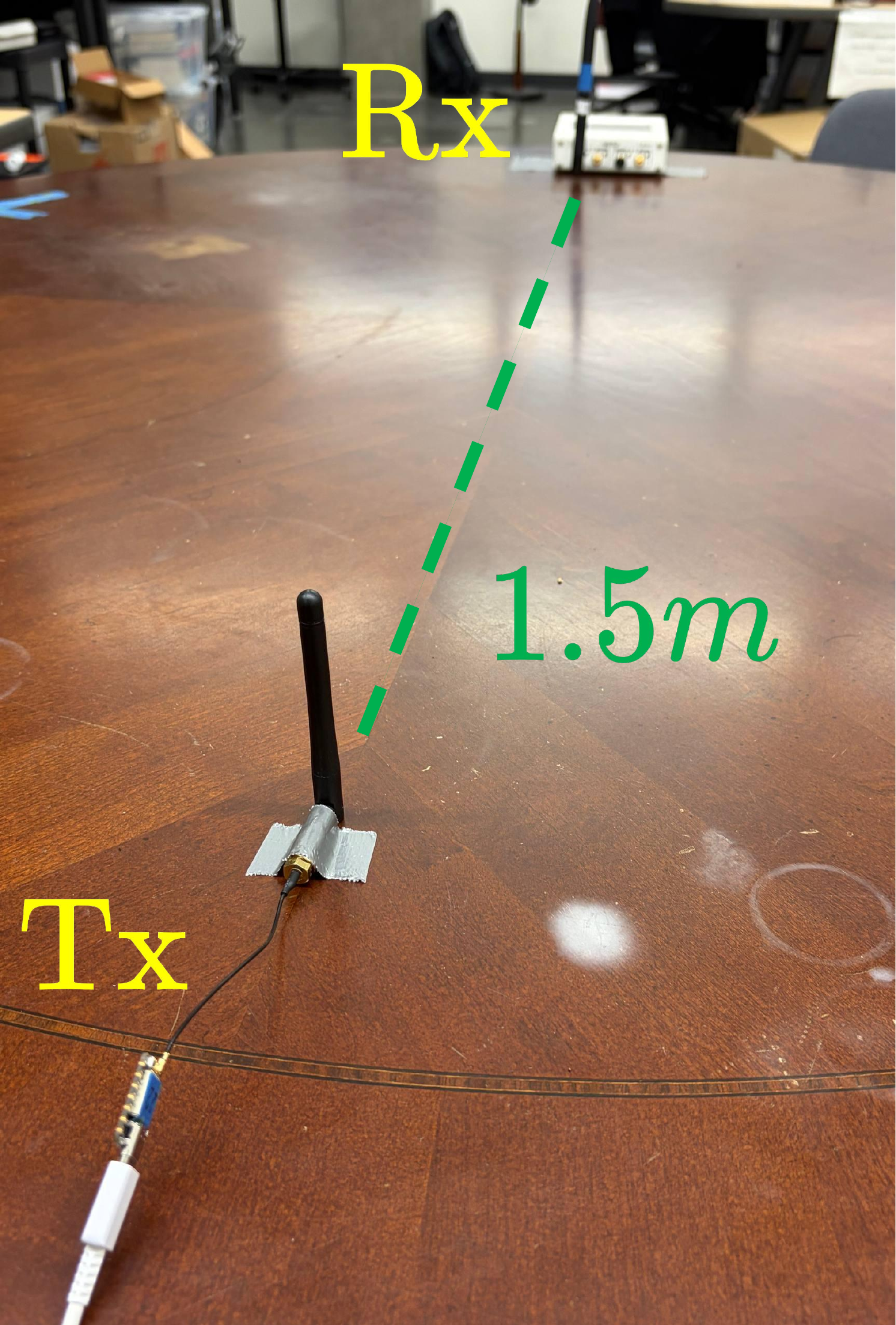}\label{subfig:1.5mLoS}}
\subfigure[1.5m NLoS]{\includegraphics[width=0.144\linewidth]{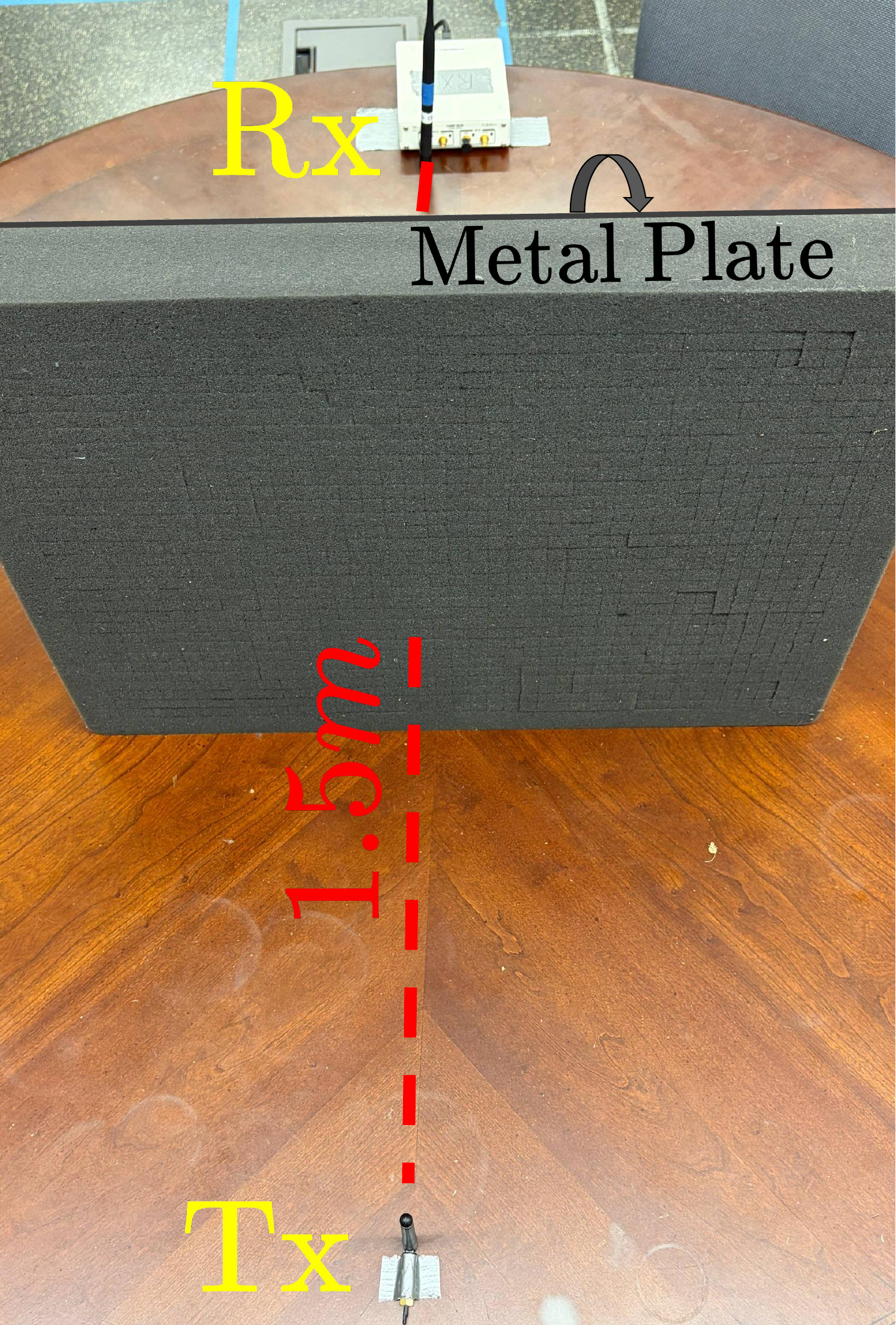}\label{subfig:1.5mNLoS}}
\subfigure[7.5m LoS]{\includegraphics[width=0.144\linewidth]{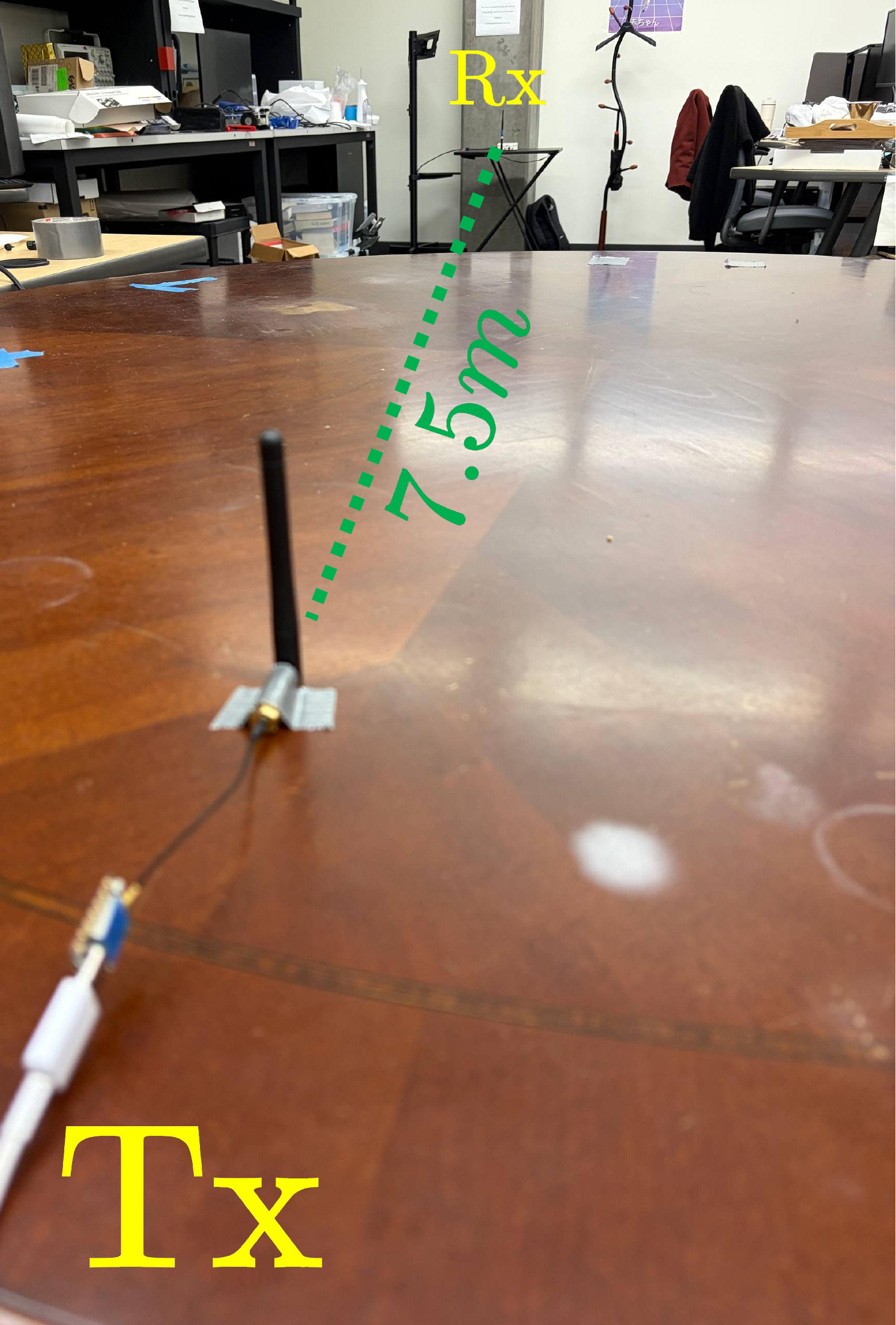}\label{subfig:7.5mLoS}}
\subfigure[Wired]{\includegraphics[width=0.1755\linewidth]{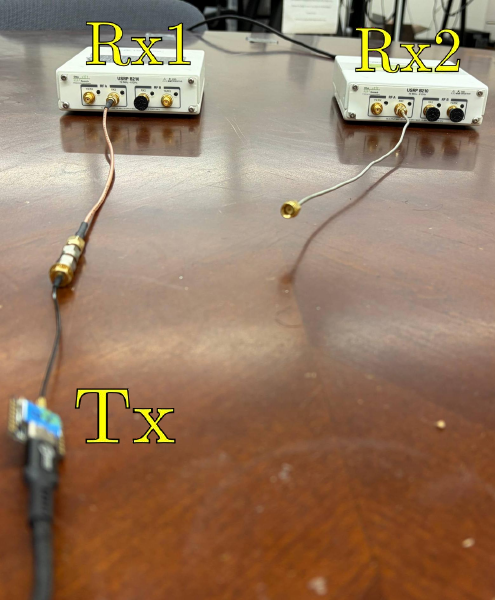}\label{subfig:Rxs}}
\subfigure[Outdoor setup (different Tx locations)]{\includegraphics[width=0.25875\linewidth]{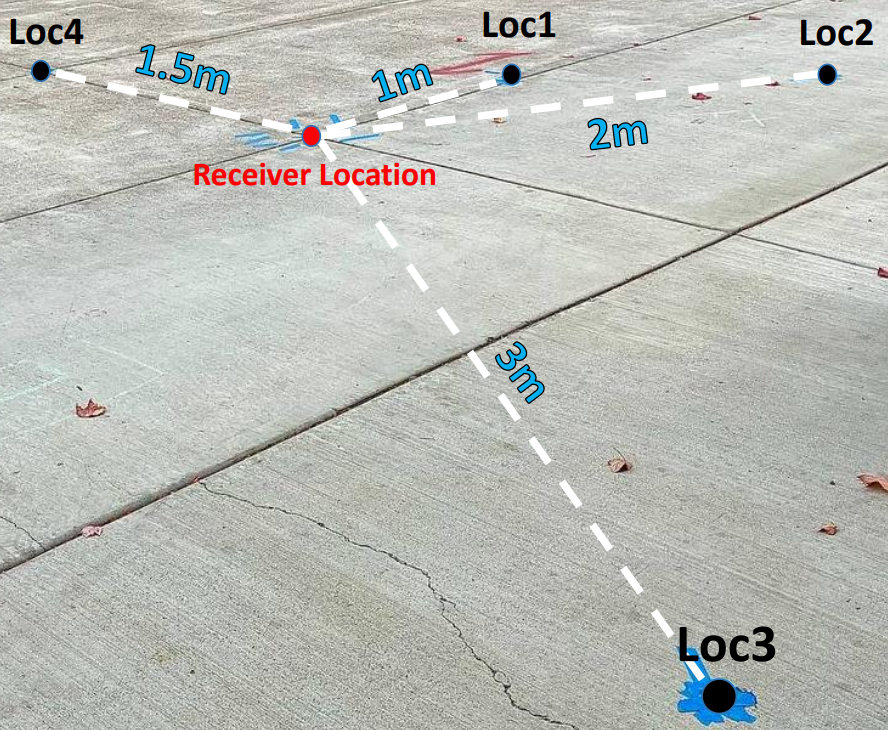}\label{subfig:OutdoorsEnv}}
    \caption{Different scenarios of data collection.}
    \label{fig:Fulltestbed}
\vspace{-10pt}
\end{figure*}


To thoroughly evaluate the resilience and domain generalization capabilities of deep learning-based RF fingerprinting models, we structured 15 data collections (D1--D15) around distinct experimental configurations. Each of these setups focuses on studying one or more domain shifts in isolation from all other domains. Table~\ref{tab:dataset_summary} summarizes these setups/datasets, and the following subsections provide a detailed explanation on how they were collected along with their potential applications. 



\subsection{Data Collection Under Varying Temperature}
Three experiments were done to investigate the dynamic behavior of IoT transmitter hardware impairments under operational thermal shifts. To track these hardware fluctuations precisely, each ESP32-C3 transmitter was configured to report its real-time internal temperature telemetry dynamically within the signal payload, as detailed in Section~\ref{sec:testbed}. To capture the full spectrum of the device's thermal evolution---from a completely unpowered state to maximum operational equilibrium---the receiver was initialized to continuously record the spectrum prior to powering on the "cold-state" IoT transmitter. Once energized, the device remained continuously operational for a duration of 20 minutes, during which it periodically broadcasted BLE packets. The true, real-time temperature measurements were embedded within the final two bytes of the packet's payload field. Figure~\ref{fig:TempvsTime_Workshop}, which shows the average device temperature vs time, suggests that devices experience warm-up during the first 6 minutes. 

Moreover, we configured the remaining payload content of size 240 Bytes to be completely random for every transmission. To analyze the joint impact of thermal drift and spatial fading, this experiment was systematically repeated across three distinct over-the-air wireless propagation channels: a 1.5m LoS link, a 1.5m NLoS link, and a 7.5m LoS link characterized by a rich multipath environment, which are shown in Figures~\ref{subfig:1.5mLoS},~\ref{subfig:1.5mNLoS} and~\ref{subfig:7.5mLoS}, respectively. All signals for this setup were captured over Ch1 using Rx2, and the associated datasets are labeled as D1, D2 and D3 in Table~\ref{tab:dataset_summary}. 


\subsection{Data Collection Under Varying Frequency Channels}
This experiment focuses only on the impact of changing the frequency channel---or equivalently the center frequency of the carrier---on the estimated device fingerprints. For that, we collected six datasets, four under a wired setup as shown in Figure~\ref{subfig:Rxs} that completely isolates the surrounding noise and interference. The four channels (Ch1, Ch2, Ch14 and Ch32) were used in this case. The other two datasets were under 1.5m LoS setup (shown in Figure~\ref{subfig:1.5mLoS}) using only Ch1 and Ch2. Since these experiments were meant to only focus on the channel impact excluding the receiver hardware and temperature variations, we only used Rx1 to capture these signals. And to mitigate the temperature dependency, we let all the devices warm-up for 6 minutes after being powered-on~\cite{albousayri2026characterizing}. 
These datasets are summarized in Table~\ref{tab:dataset_summary}.


\subsection{Data Collection Across Different Receivers}
To study the impact of receiver hardware, we collected signals using both Rx1 and Rx2. We collected one dataset under the wired connection (as shown in Figure~\ref{subfig:Rxs}) and another under the wireless 1.5m LoS setup (shown in Figure~\ref{subfig:1.5mLoS}). Signal collection was performed post warm-up (after a 6-minute wait) using only Ch1. The cross-receiver wired datasets are labeled D7 and D8, while the wireless datasets are D10 (corresponding to Rx1) as well as D11-a and D11-b (corresponding to Rx2 collected across two different days).


\subsection{Data Collection Under Varying Environments/Locations}
Using only Rx1 and Ch1 and after device warm-up, we collected data on wireless outdoors setup across 4 different locations while keeping the receiver fixed as seen in Figure~\ref{subfig:OutdoorsEnv}. These locations have the following Tx--Rx distances respectively: 1m, 2m, 3m, 1.5m, and their associated datasets are labeled as D12, D13, D14 and D15.

\begin{table*}[t]
\centering
\caption{Summary of the proposed BLE RF fingerprinting datasets.}
\label{tab:dataset_summary}
\setlength{\tabcolsep}{5pt}
\renewcommand{\arraystretch}{0.7}

\begin{tabular}{lccccccccccc}
\toprule
\multirow{2}{*}{\textbf{Dataset}} &
\multirow{2.5}{*}{\makecell{\textbf{\# Frames}\\\textbf{/Device}}} &
\multirow{2}{*}{\makecell{\textbf{\# Samples}\\\textbf{/Frame}}} &
\multirow{2.5}{*}{\textbf{Payload}} &
\multirow{2.5}{*}{\makecell{\textbf{Stable/}\\\textbf{Warming-up}}} &
\multirow{2.5}{*}{\makecell{\textbf{Temperature}\\\textbf{Included}}} &
\multicolumn{6}{c}{\textbf{Collection Conditions}} \\
\cmidrule(lr){7-12}

&
&
&
&
&
&
\textbf{\scriptsize Medium} &
\textbf{\scriptsize Environment} &
\textbf{\scriptsize Channel} &
\textbf{\scriptsize Receiver} &
\textbf{\scriptsize Location} &
\textbf{\scriptsize SNR (Avg)} \\

\midrule
D1 &
33,000 &
4,096 &
Random &
Warming-up &
$\checkmark$ &
Wireless &
Indoors &
Ch1 &
Rx2 &
1.5m L & 
28--30dB \\ 

D2 &
33,000 &
4,096 &
Random &
Warming-up &
$\checkmark$ &
Wireless &
Indoors &
Ch1 &
Rx2 &
1.5m NL & 
18--22dB
\\

D3 &
33,000 &
4,096 &
Random &
Warming-up &
$\checkmark$ &
Wireless &
Indoors &
Ch1 &
Rx2 &
7.5m L &
16--19dB \\ 

\midrule

D4 &
3,000 &
1,664 &
Fixed &
Stable (6min) &
$\times$ &
Wired &
Indoors &
Ch1 &
Rx1 &
NA & 
\multirow{5}{*}{{40--50dB}} \\

D5 &
3,000 &
1,664 &
Fixed &
Stable (6min) &
$\times$ &
Wired &
Indoors &
Ch2 &
Rx1 &
NA & 
 \\

D6 &
3,000 &
1,664 &
Fixed &
Stable (6min) &
$\times$ &
Wired &
Indoors &
Ch14 &
Rx1 &
NA & 
 \\

D7 &
3,000 &
1,664 &
Fixed &
Stable (6min) &
$\times$ &
Wired &
Indoors &
Ch32 &
Rx1 &
NA & 
 \\ 

D8 &
3,000 &
1,664 &
Fixed &
Stable (6min) &
$\times$ &
Wired &
Indoors &
Ch1 &
Rx2 &
NA & 
 \\ 

\midrule
D9 &
3,000 &
1,664 &
Fixed &
Stable (6min) &
$\times$ &
Wireless &
Indoors &
Ch1 &
Rx1 &
1.5m L & 
\multirow{4}{*}{{28--30dB}} \\ 

D10 &
3,000 &
1,664 &
Fixed &
Stable (6min) &
$\times$ &
Wireless &
Indoors &
Ch2 &
Rx1 &
1.5m L & 
\\

D11-a &
3,000 &
1,664 &
Fixed &
Stable (6min) &
$\times$ &
Wireless &
Indoors &
Ch1 &
Rx2 &
1.5m L & 
\\

{D11-b} &
3,000 &
1,664 &
Fixed &
Stable (6min) &
$\times$ &
Wireless &
Indoors &
Ch1 &
Rx2 &
1.5m L & 
\\

\midrule

D12 &
3,000 &
1,664 &
Fixed &
Stable (10min) &
$\times$ &
Wireless &
Outdoors &
Ch1 &
Rx1 &
1m L &
\multirow{4}{*}{{30--35dB}} \\

D13 &
3,000 &
1,664 &
Fixed &
Stable (10min) &
$\times$ &
Wireless &
Outdoors &
Ch1 &
Rx1 &
2m L &
\\  

D14 &
3,000 &
1,664 &
Fixed &
Stable (10min) &
$\times$ &
Wireless &
Outdoors &
Ch1 &
Rx1 &
3m L &
\\  

D15 &
3,000 &
1,664 &
Fixed &
Stable (10min) &
$\times$ &
Wireless &
Outdoors &
Ch1 &
Rx1 &
1.5m L &
\\  
\bottomrule
\end{tabular}
\end{table*}

\section{Dataset Format and Description}
\label{sec:dataset}
Captured baseband signals were preprocessed to remove the silence period and synchronize the signals. Packet synchronization was performed to align each signal by finding the optimal arrival index $n_{start}$ using cross-correlation against the known BLE preamble:
$n_{start} = \arg\max_{n} \left\vert{} \sum_{k=0}^{L-1} r[n+k] \cdot p^*[k] \right\vert{}$, where $r[n]$ is the raw received complex stream, $p[k]$ is the local preamble reference sequence of length $L$, and $(\cdot)^*$ denotes the complex conjugate.
Following alignment, we kept the transient period of the BLE frame of size $N=91$ samples (prior to the preamble).
Retaining this region enables the utilization of transient-state and steady-state hardware characteristics, which is particularly used for RF fingerprinting tasks~\cite{uzundurukan2020database}.

The dataset is structured hierarchically by device cohort. For each device $k \in \{1, 2, \dots, 31\}$, the captured transmissions are organized into a uniform data matrix $\mathbf{X}_k \in \mathbb{C}^{D \times N}$, where $D$ represents the total number of recorded packet bursts (frames) and $N$ denotes the number of complex IQ samples per packet, stored as complex64 type. 
For experimental scenarios involving thermal measurements, each device profile additionally includes a synchronized temperature metadata vector $\mathbf{T}_k \in \mathbb{R}^{D}$, where each element represents the real-time floating-point (float32 type) temperature measurement reported at the time of transmission.
To ensure seamless integration with modern deep learning pipelines, both $\mathbf{X}_k$ and $\mathbf{T}_k$ are stored and distributed as native NumPy arrays.

\begin{figure}
    \centering
    \includegraphics[width=\linewidth]{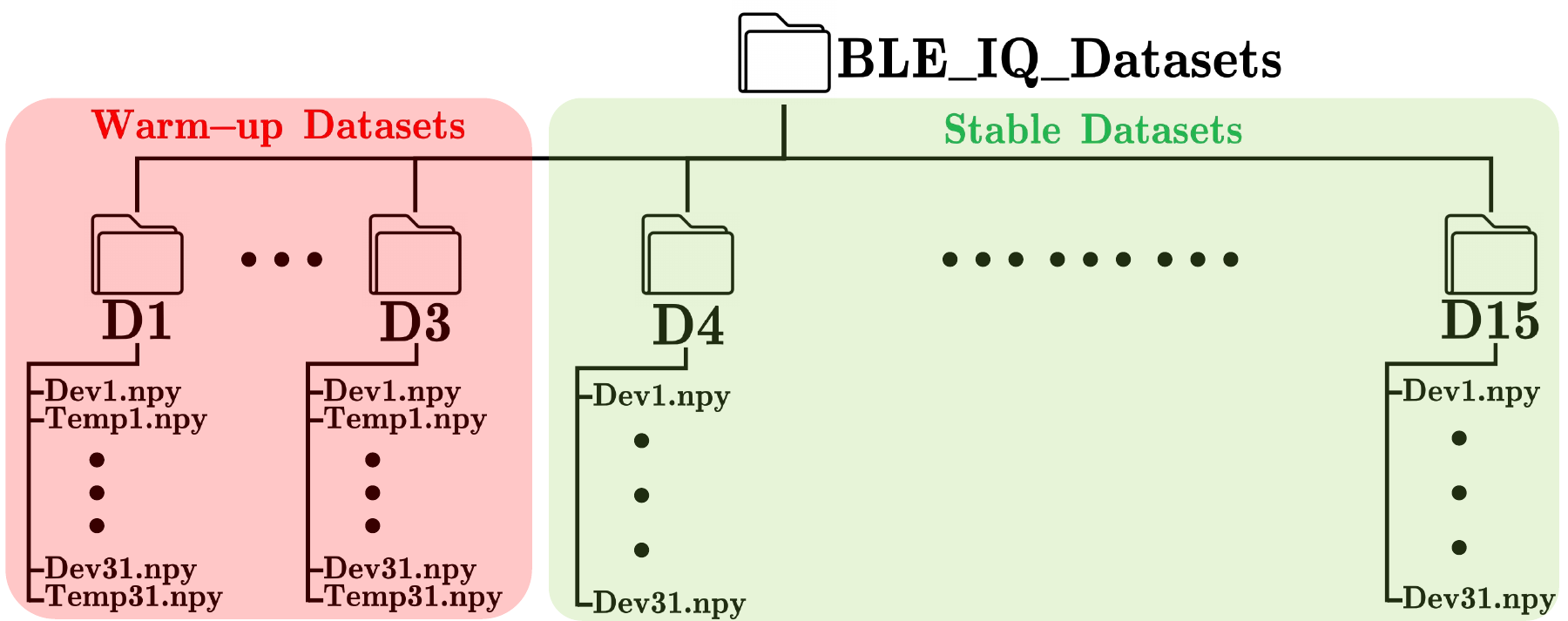}
    \caption{Online datasets organizations.}
    \label{fig:DirectoryTree}
\end{figure}

\section{Use Case: Device Identification via RFFP}
\label{sec:applications}


\begin{table*}[t]
    \centering
    \caption{Average identification accuracy (\%) for 31 devices (classes) of different NN architectures using different feature representations across testing domains, when the training domain is D12.}
    \label{tab:domain_feature_nn}
    \scriptsize
    \setlength{\tabcolsep}{2pt}
    \renewcommand{\arraystretch}{0.9}
    \begin{tabular}{c|cccc|cccc|cccc|cccc|cccc|cccc}
        \toprule
        \multirow{2}{*}{%
          \raisebox{0.5\height}{%
            \diagbox[width=6em,height=3em]
              {\rotatebox{0}{\textbf{Domain}}}
              {\raisebox{5pt}{\rotatebox{0}{\textbf{Technique}}}}%
          }%
        }
                
        & \multicolumn{4}{c|}{\textbf{IQ}}
        & \multicolumn{4}{c|}{\textbf{PD~\cite{albousayri2025bluetooth}}}
        & \multicolumn{4}{c|}{\textbf{CLPS~\cite{tang2024causal}}}
        & \multicolumn{4}{c|}{\textbf{DoLoS~\cite{xing2022design}}}
        & \multicolumn{4}{c|}{\textbf{Mbed~\cite{jagannath2023embedding}}}
        & \multicolumn{4}{c}{\textbf{EPS~\cite{elmaghbub2023eps}}} \\
        
        \cmidrule(lr){2-5}
        \cmidrule(lr){6-9}
        \cmidrule(lr){10-13}
        \cmidrule(lr){14-17}
        \cmidrule(lr){18-21}
        \cmidrule(lr){22-25}

        & CNN & GRU & LSTM & TR.
        & CNN & GRU & LSTM & TR.
        & CNN & GRU & LSTM & TR.
        & CNN & GRU & LSTM & TR.
        & CNN & GRU & LSTM & TR.
        & CNN & GRU & LSTM & TR. \\

        \midrule
        
        \textbf{D12} & 99.7 & 99.7 & 99.8 & 99.7 & 99.8 & 99.9 & 99.8 & \textbf{99.9} & 99.8 & 99.8 & 99.8 & 99.8 & 99.8 & 99.8 & 99.8 & 99.9 & 99.3 & 99.5 & 99.2 & 99.6 & 98.8 & 98.8 & 98.9 & 98.7 \\
        \textbf{D13} & 96.4 & 95.8 & 95.4 & 95.6 & 98.1 & 98.6 & \textbf{98.6} & 98.5 & 92.1 & 90.8 & 91.6 & 91.2 & 95.2 & 93.7 & 94.8 & 95.1 & 84.5 & 86.7 & 84.8 & 86.8 & 80.8 & 79.8 & 81.1 & 79.4 \\
        \textbf{D14} & 77.9 & 77.0 & 73.4 & 75.5 & 90.7 & 90.5 & \textbf{90.9} & 90.2 & 57.5 & 58.4 & 60.3 & 56.9 & 66.3 & 67.9 & 66.5 & 67.2 & 53.3 & 51.9 & 51.1 & 52.9 & 48.7 & 49.1 & 49.3 & 48.1 \\
        \textbf{D15} & 82.8 & 82.8 & 82.1 & 82.1 & \textbf{95.6} & 95.2 & 94.5 & 94.8 & 76.8 & 77.4 & 78.4 & 78.1 & 79.5 & 79.2 & 79.2 & 79.1 & 70.1 & 70.5 & 69.2 & 71.9 & 65.0 & 64.5 & 64.3 & 65.1 \\

        \bottomrule
    \end{tabular}
    \vspace{-16pt}
\end{table*}

In this section, we use the collected datasets to assess the effectiveness of different RF fingerprinting (RFFP) feature extraction techniques and investigate their ability to generalize to changes in receiver hardware and transmitter location.
To study the impact of location changes, we train our model on data collected at one location (D12) and  evaluate its device identification performance on data collected across four different locations (D12--D15). Beyond spatial generalization, we also investigated how RFFP models overfit to the receiver hardware used to collect the training data; for this, we trained the model on data collected using Rx1 indoor (D9) and tested it on data collected using Rx2 indoor (D11-a), and vice versa.

The RFFP feature extraction techniques used in this evaluation can be grouped into signal processing-based and DL-based. For the signal processing-based techniques, we investigated the following approaches: \textbf{IQ~\cite{kashani2024radio}}, \textbf{PD~\cite{albousayri2025bluetooth}}, \textbf{CLPS~\cite{tang2024causal}}, \textbf{DoLoS~\cite{xing2022design}}, \textbf{Mbed~\cite{jagannath2023embedding}} and \textbf{EPS~\cite{elmaghbub2023eps}}. For the DL-based techniques, these neural network (NN) architectures are evaluated: \textbf{Convolutional Neural Network (CNN)}, \textbf{Gated Recurrent Unit (GRU)}, \textbf{Long Short-Term Memory (LSTM)}, and \textbf{Transformer (TR.)}, following the same structure and hyper-parameters in~\cite{albousayri2026characterizing}. All $K\!=\!31$ devices (classes) and only the first $N'=512$ time samples (out of $N\!\!=\!\!1664$) were utilized for classification. All results are reported under 3 random seeds.

Table~\ref{tab:domain_feature_nn} benchmarks the identification accuracy---defined as the total number of correctly classified signals by the total number of signals---for every combination of signal processing features and NN architectures.
The table highlights that across all baselines, the choice of NN architecture tends to have less impact on the performance, offering only about $\pm 2\%$ accuracy gain.
Another key observation is that across all baselines, testing under the same training domain (D12) achieves high accuracy; however, the performance degrades when testing is done over unseen domains.
Averaging across these unseen domains and NN architectures, \textbf{PD} achieves the state of the art performance of $94.68\%$, \textbf{IQ} comes second with $84.73\%$, then \textbf{DoLoS} with $80.32\%$, then \textbf{CLPS} with $75.78\%$,  then \textbf{Mbed} with $69.46\%$, and finally \textbf{EPS} with $64.61\%$.


\begin{figure}
    \centering
    \includegraphics[width=\linewidth]{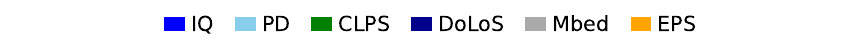}
    \vspace{-16pt}

    \subfigure[Trained on D9]{\includegraphics[width=0.49\linewidth]{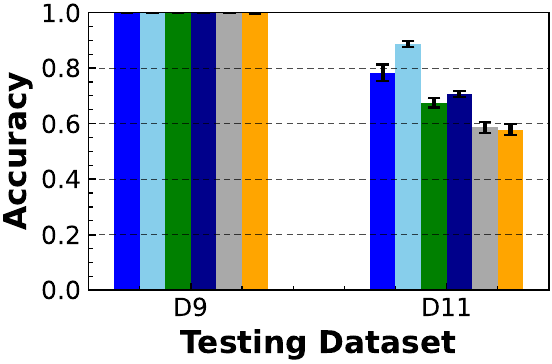}\label{subfig:RxTrainedonR1}}
    \subfigure[Trained on D11-a]{\includegraphics[width=0.49\linewidth]{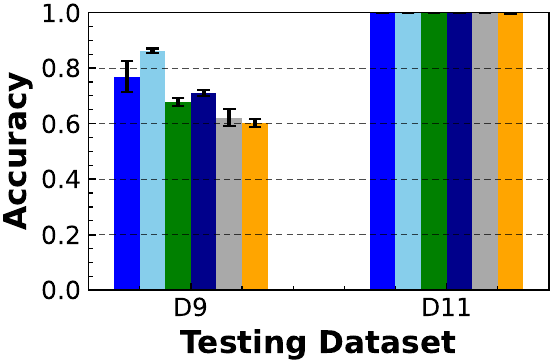}\label{subfig:RxTrainedonR2}}
    \caption{Cross-Receivers benchmark evaluation.}
    \label{fig:Cross-Receivers}
\end{figure}

Figures~\ref{fig:Cross-Receivers}(a-b) show the identification accuracy when the model is trained and cross tested on different receivers (i.e., using D9 and D11-a datasets), averaged across all seeds and NN setups. These results support our earlier findings on the generalization of \textbf{PD}, showing that the other techniques maintain the same performance order across unseen receivers.


In addition to the study of RF fingerprinting's ability to adapt to location changes and unseen receiver hardware, the datasets provide a benchmark for evaluating RFFP under other possible domain changes, including:
\begin{enumerate}[leftmargin=*]
\item {\bf {Changes in thermal state.}}
By utilizing datasets D1--D3 (which include the synchronized temperature vector $\mathbf{T}_k$ across a continuous warming-up curve), researchers can benchmark how severe thermal drift causes feature space misalignment. Models can be trained on stable, room-temperature snapshots (D9) and cross-tested against active warming-up phases (D1) to evaluate classifier resilience to internal hardware heating. 

\item {\bf {Changes in transmission environment.}}
Evaluating model resilience under severe propagation environment shifts. For example, a classifier can be trained on indoor data (D9) and evaluated directly against outdoor open-field captures (D12--D15).


\item {\bf {Changes in wireless channel.}} 
Investigating model stability across varying BLE frequencies. Researchers can optimize classifiers using a single channel (D4 on Ch1) and cross-test them across other channels (D5, D6, D7 on Ch2, Ch14, Ch32) to check for channel-invariant feature extraction.

\item {\bf {Short- vs long-term fingerprint aging.}}
Quantifying fingerprint decay and drift over explicit time intervals. Short-term aging can be analyzed by cross-testing between the day1 and day2 of D11, while long-term aging can be evaluated by comparing D8 and D11 against the D1--D3 where there is around a year gap between the two sets.

\item {\bf {Fine-grained open-set identification.}}
Evaluates RFFP systems under open-set authentication scenarios where unauthorized or rogue devices attempt to impersonate legitimate transmitters~\cite{albousayri2026replicating}.

\end{enumerate}

\section{Beyond RFFP: Other Use Cases}
\label{sec:applications2}
\subsection{Hardware Impairment Estimation}
This dataset can also serve as a diagnostic utility for characterizing, modeling, and compensating for hardware-level distortions to optimize wideband receiver performance and link reliability. By pairing the complex IQ signal matrices with the synchronized temperature telemetry vectors ($\mathbf{T}_k$) across the active warming-up curves in D1--D3, researchers can track how internal oscillator behavior, power amplifier responses, and IQ imbalances fluctuate dynamically as a function of thermal drift. This multi-domain data enables the design of advanced regression and calibration algorithms capable of tracking thermal frequency pulling and constellation warping. Furthermore, evaluating these imperfections across distinct RF channels (D4--D7) and dual-receiver architectures (D8) allows for comprehensive cross-channel consistency analysis, enabling the development of robust channel equalization techniques that preserve communication integrity under harsh operational and hardware stresses.

\subsection{Integrated Sensing and Communications (ISAC)}
By integrating systematic spatial and environmental configurations alongside standard communication packets, this dataset offers a valuable framework for exploring joint location-sensing and data decoding. The structural inclusion of indoor LoS propagation (D1), indoor non-LoS multipath conditions (D2, D3), and outdoor variable-distance tracks (D12--D15) provides the exact physical-layer channel diversity required to train machine learning models for environment profiling and link-type classification. 


\section{Conclusion}
\label{sec:conclusion}
This paper presents a comprehensive BLE RF fingerprinting dataset spanning multiple experimental scenarios, specifically designed to enable thorough performance evaluations of AI-driven wireless techniques. This dataset is made publicly available to the research community to serve as a robust benchmark for developing RF domain adaptation frameworks.


\bibliographystyle{IEEEtran}
\bibliography{References}

@inproceedings{elmaghbub2021comprehensive,
  title={Comprehensive RF dataset collection and release: A deep learning-based device fingerprinting use case},
  author={Elmaghbub, Abdurrahman and Hamdaoui, Bechir},
  booktitle={2021 IEEE Globecom Workshops (GC Wkshps)},
  pages={1--7},
  year={2021},
  organization={IEEE}
}

@article{tang2024causal,
  title={Causal learning for robust specific emitter identification over unknown channel statistics},
  author={Tang, Peng and Ding, Guoru and Xu, Yitao and Jiao, Yutao and Song, Yehui and Wei, Guofeng},
  journal={IEEE Transactions on Information Forensics and Security},
  volume={19},
  pages={5316--5329},
  year={2024},
  publisher={IEEE}
}

@article{xing2022design,
  title={Design of a channel robust radio frequency fingerprint identification scheme},
  author={Xing, Yuexiu and Hu, Aiqun and Zhang, Junqing and Peng, Linning and Wang, Xianbin},
  journal={IEEE Internet of Things Journal},
  volume={10},
  number={8},
  pages={6946--6959},
  year={2022},
  publisher={IEEE}
}

@inproceedings{albousayri2025bluetooth,
  title={Bluetooth Fingerprint Identification Under Domain Shift Through Transient Phase Derivative},
  author={Albousayri, Haytham and Hamdaoui, Bechir and Wong, Weng-Keen and Basha, Nora},
  booktitle={2025 IEEE Conference on Communications and Network Security (CNS)},
  pages={1--9},
  year={2025},
  organization={IEEE}
}

@inproceedings{albousayri2025neural,
  title={Neural Network-Driven Estimation of Hardware Impairments for Robust Wireless Device Identification},
  author={Albousayri, Haytham and Hamdaoui, Bechir and Wong, Weng-Keen},
  booktitle={ICML 2025 Workshop on Machine Learning for Wireless Communication and Networks (ML4Wireless)}
}

@techreport{bluetooth60,
  title        = {Bluetooth Core Specification Version 6.0},
  institution  = {Bluetooth Special Interest Group},
  year         = {2023},
  url          = {https://www.bluetooth.com/specifications/bluetooth-core-specification/},
  note         = {Accessed: February 03, 2025}
}

@inproceedings{kashani2024radio,
  title={Radio Frequency Fingerprinting in WBANs Using Complex-Valued Convolutional Neural Networks},
  author={Kashani, SeyedMohammad and Sherazi, Syed Mujtaba and Khokhar, Ashfaq and Kim, Sang Wu and Nait-Abdesselam, Farid},
  booktitle={2024 International Wireless Communications and Mobile Computing (IWCMC)},
  year={2024},
  organization={IEEE}
}

@article{jagannath2023embedding,
  title={Embedding-assisted attentional deep learning for real-world RF fingerprinting of Bluetooth},
  author={Jagannath, Anu and Jagannath, Jithin},
  journal={IEEE Transactions on Cognitive Communications and Networking},
  year={2023},
  publisher={IEEE}
}

@article{elmaghbub2023eps,
  title={EPS: distinguishable IQ data representation for domain-adaptation learning of device fingerprints},
  author={Elmaghbub, Abdurrahman and Hamdaoui, Bechir},
  journal={arXiv preprint arXiv:2308.04467},
  year={2023}
}

@article{johnson2025domain,
  title={Domain-adaptive device fingerprints for network access authentication through multifractal dimension representation},
  author={Johnson, Benjamin and Albousayri, Haytham and Hamdaoui, Bechir and Dunn, Lucas},
  journal={IEEE Network},
  year={2025},
  publisher={IEEE}
}

@article{monteiro2026deep,
  title={Deep Learning for Bluetooth Low Energy Indoor Positioning: A Comparative Analysis of Fingerprinting Accuracy and Generalization to Unseen Locations},
  author={Monteiro, Marcos Eduardo Pivaro and Dos Santos, Andr{\'e} Badenas and Peron, Guilherme De Santi and Rayel, Ohara Kerusauskas and Dos Santos, Eduardo Nunes},
  journal={IEEE Access},
  volume={14},
  year={2026},
  publisher={IEEE}
}

@data{1s6a-tv49-26,
doi = {10.21227/1s6a-tv49},
url = {https://dx.doi.org/10.21227/1s6a-tv49},
author = {Igor Pereira and Guilherme Peron and Marcos Monteiro and Glauber Brante and Ohara Rayel and Guilherme Moritz and Richard Souza},
publisher = {IEEE Dataport},
title = {BLE 5.1 Pose Dataset},
year = {2026} }

@article{uzundurukan2020database,
  title={A database for the radio frequency fingerprinting of Bluetooth devices},
  author={Uzundurukan, Emre and Dalveren, Yaser and Kara, Ali},
  journal={Data},
  pages={55},
  year={2020},
  publisher={MDPI}
}

@inproceedings{albousayri2026characterizing,
  title={Characterizing and Mitigating the Effects of Device Temperature on RF Fingerprinting Accuracy},
  author={Albousayri, Haytham and Hamdaoui, Bechir},
  booktitle={2026 17th International Conference on Network of the Future (NoF)},
  year={2026},
  organization={IEEE}
}

@inproceedings{albousayri2026replicating,
  title={Replicating the Signature: Unsupervised Targeted Impersonation Attack on RF Fingerprinting},
  author={Albousayri, Haytham and Hamdaoui, Bechir},
  booktitle={2026 23rd Annual International Conference on Privacy, Security and Trust (PST)},
  pages={1--10},
  year={2026},
  organization={IEEE}
}

@article{hanna2022wisig,
  title={WiSig: A large-scale WiFi signal dataset for receiver and channel agnostic RF fingerprinting},
  author={Hanna, Samer and Karunaratne, Samurdhi and Cabric, Danijela},
  journal={IEEE Access},
  volume={10},
  pages={22808--22818},
  year={2022},
  publisher={IEEE}
}

@String{Computing = "Computing" }

@ArtifactSoftware{R,
    title = {R: A Language and Environment for Statistical Computing},
    author = {{R Core Team}},
    organization = {R Foundation for Statistical Computing},
    address = {Vienna, Austria},
    year = {2019},
    url = {https://www.R-project.org/},
}

\end{document}